\documentstyle[12pt,aasms4]{article}

\righthead{Magnetic Field Generation in Galaxy Clusters}
\lefthead{V\"olk \& Atoyan}

\newcommand{\gsim}{\,\raisebox{0.2em}{$>$}\!\!\!\!\!
\raisebox{-0.25em}{$\sim$}\,}
\newcommand{\lsim}{\,\raisebox{0.2em}{$<$}\!\!\!\!\!
\raisebox{-0.25em}{$\sim$}\,}
\begin{document}
\title{Early Starbursts and Magnetic Field Generation \\
in Galaxy Clusters}
\author{H. J. V\"olk} 

\affil{Max--Planck--Institut f\"ur Kernphysik, P.O. Box 103980, 
D-69029 Heidelberg, Germany}
 \authoremail{Heinrich.Voelk@mpi-hd.mpg.de}

\and 

\author{A. M. Atoyan}

\affil{Max--Planck--Institut f\"ur Kernphysik, P.O. Box 103980, 
D-69029 Heidelberg, Germany}
\affil{Yerevan Physics Institute, Alikhanian Broth. 2, 375036 Yerevan,
Armenia}
\authoremail{Armen.Atoyan@mpi-hd.mpg.de}

\begin{abstract}
We propose a mechanism for the early generation of the mean
Intracluster magnetic field in terms of magnetized galactic winds.
These winds are the result of starburst phases of the cluster
galaxies, assumed to produce the predominant population of early type
galaxies in mergers of gas-rich progenitors. After further cluster
contraction typical field strengths are $10^{-7}$ Gauss. This
estimate may increase to the level of $10^{-6}$ Gauss if more
extreme galactic parameters, and subsequent shear amplification of
the field are considered. The topology of the field is one of almost
unconnected Wind Bubbles with Parker-type spiral field configurations
over scales of the distance between galaxies. Further cluster
accretion, that continues chaotically in space and time up to the
present, will perturb these "large-scale" mean fields on smaller or
at best comparable spatial scales. The small scale fields in the
resulting turbulent fluctuation spectrum should be able to confine
relativistic particles over times longer than the age of the
Universe. The nonthermal particle content of galaxy clusters should
therefore also have a "cosmological" hadronic component generated
during the early starburst phase of the member galaxies. Already by
itself it implies a nonthermal energy fraction of about 10 percent
for the Intracluster gas which should then be detectable by future
gamma-ray telescopes.
\end{abstract}


\section{Introduction}

Rich clusters of galaxies are the largest gravitationally bound structures
in the Universe and should confine a representative fraction of its mass.
Therefore the study of their dynamical properties and radiation content
should allow, amongst other things, interesting cosmological conclusions
on the relative amounts of visible and dark baryonic matter, and of
nonbaryonic matter (\cite{whifab95,tur99}).

Another basic characteristic, due to energetic particle confinement,
is the ratio of nonthermal to thermal energy in these objects. To a
significant extent that ratio should be pre-determined during the
epoch of early starburst activity and thus preserve the energetic
history of cluster formation. The necessary confinement of the
nonthermal particle components is intimately related to the existence
of strong and chaotic magnetic fields in the Intracluster Medium
(ICM), and we shall propose physical mechanisms for their early
generation as well as for their present fluctuations.

In principle, detailed ab initio simulations of the dynamics of cluster
formation under the dominant gravitational influence of the dark matter
component (\cite{kau99}) should establish the overall cosmological
framework for the present considerations. We rather start in a
complementary way with the discussion of a simplified model of cluster
formation and of chemical enrichment of the Intracluster gas. It has the
advantage that it directly allows a discussion of the
physical processes of nonthermal particle production and confinement. The
main part of the paper concerns a proposal of cluster magnetic field
generation in terms of galactic winds due to early starbursts and their
amplification effect on magnetic fields drawn out from the progenitors of
today's cluster galaxies into Intracluster space. It is argued that due to
these dynamical processes there is no need for the operation of a
dissipative turbulent dynamo in the ICM. The ongoing cluster accretion
naturally leads to a strong fluctuating part of the Intracluster magnetic
fields. A detailed discussion of the nonthermal radiation from galaxy
clusters will be given in a separate paper (\cite{atovol2000}).

\section{Rich Clusters}

We shall be concerned here with rich clusters, i.e. conglomerates
with typically more than 100 member galaxies. They have typical radii
$R_{\rm cl} \sim$ few Mpc and baryonic masses $M_{\rm cl} \sim
10^{14} \,{\rm to}\, 10^{15}\,{\rm M}_{\odot}$. Many such clusters
are rather evolved and contain predominantly early type S0 and
E-galaxies, at least in their inner parts. Examples for bright and
relatively nearby clusters of this type are the Perseus and the Coma
clusters with distances $d \sim 100\,\rm Mpc$. The Perseus cluster is
the brightest cluster in soft X-rays.  The large X-ray luminosity is
due to the very hot ($T \sim 10^7 \,\rm to \, 10^8\,$K), massive
($M_{\rm gas} \sim {\rm few} \times \sum M_{\rm gal}$), and
metal-rich ($[{\rm Fe}]_{\rm cl} \simeq 0.35 [{\rm Fe}]_{\odot}$) ICM
gas (\cite{boh96}). As a consequence the gas pressures are extremely high,
with $nT$ ranging from $10^3\, {\rm to}\, 10^5\, {\rm K cm}^{-3}$.

\subsection{Cluster Formation} 

The metallicity of the ICM gas, for instance in terms of the fractional
ICM iron mass, is correlated with the total optical luminosity in the E
and S0 galaxies of rich clusters (\cite{arn92}).  The correlation supports
the view that early starbursts due to galaxy-galaxy interactions of
gas-rich progenitors have produced a large number of core collapse
Supernovae due to massive stars (for simplicity referred to here as SNe).
They should have heated the originally present interstellar gas and
generated violent galactic winds which removed the Interstellar Medium,
leaving gas-poor E and S0 galaxies behind. This mass loss should have led
to the observed strong chemical enrichment of the ICM gas. We also
conjecture that the ionizing radiation, the winds, and the large-scale
shocks from these early galaxy mergers - together with the hard radiation
from AGNs - strongly heated the remaining primordial ICM gas, and thus
prevented further galaxy formation.  A quantitative discussion of the
dynamical prerequisites for galactic winds and of the total number of SNe
in clusters is given by V\"olk et al. (\cite[hereafter referred to as
Paper I]{paper1}).

The total number of SNe since galaxy formation in the cluster, roughly a
Hubble time $T_{\rm H} \simeq 1.5 \times 10^{10} {\rm yr}$ ago, is then
given by $$N_{\rm SN}=\int_{-T_{\rm H}}^{0}dt \times \nu_{\rm
SN}(t)=\frac{0.35 \,{[Fe]}_{\odot} \times M_{\rm cl}}{\delta
M_{\rm Fe}}\;,$$
where $\delta M_{\rm Fe}$ is the amount of iron produced per event. In
such
starbursts we dominantly expect core collapse SNe from massive progenitor
stars to occur, with $\delta M_{\rm Fe}\simeq 0.1 M_{\odot}$ on average.
For
the Perseus cluster this implies $N_{\rm SN}^{\rm Perseus} \sim 3 \times
10^{12}$. The corresponding total energy input into the interstellar
medium is $N_{\rm SN}E_{\rm SN} \sim 3 \times 10^{63} E_{51}\,\rm erg$,
where $E_{51}=(E_{\rm SN}/10^{51} \,\rm erg)$ is the average hydrodynamic
energy release per SN in units of $10^{51}\,\rm ergs$.

Assuming the early starbursts to occur at a typical redshift of $z\sim 2$ 
due to the merging of gas-rich progenitors in an overdense protocluster
environment (\cite{ste93}), with a duration of 
$T_{\rm SB} \lsim 10^9 \,\rm yr$, we obtain
$$
\frac{(N_{\rm SN}^{\rm Perseus}/N_{\rm gal}^{\rm Perseus})}{T_{\rm SB}}
\gsim 100 \times \nu_{\rm SN}^{\rm Milky\, Way} ,
$$
where $\nu_{\rm SN}^{\rm Milky\, Way}$ is taken as 1/(30\,yr), and 
$N_{\rm gal}^{\rm Perseus} \!\simeq \! 500$ denotes the number of galaxies 
in the Perseus
cluster. As an example we can compare to the archetypical contemporary
starburst galaxy $M82$. It has a current SN rate $\nu_{\rm SN}^{\rm M82}
\sim
10 \times \nu_{\rm SN}^{\rm Milky\, Way}$, a wind 
velocity $v_{\rm wind} \sim 2300\, {\rm
km/sec}$, and a mass-loss rate of $\dot{M} \sim 0.8 M_{\odot}/{\rm yr}$
(\cite{bre94}). The starburst nucleus of M82 is characterized
by the following values for the interstellar gas temperature $T$, gas
density $n$, and thermal gas pressure $p$ at the base of the wind: 
$T_{\rm base} \sim 10^8 \,{\rm K}$,
$n_{\rm base} \sim 0.3\, {\rm cm}^{-3}$, and 
$p_{\rm base}/k_{\rm B} \sim 10^7\, {\rm K \,cm^{-3}}$ (\cite{sch89}). 
Since the thermal ICM gas pressure in the Perseus cluster is
$p_{\rm cl}^{\rm Perseus}/k_{\rm B} \sim 10^4 \,{\rm K}\,
{\rm cm}^{-3}$, it is clear 
that an object like M82 could readily drive a wind even against the 
{\it present-day} ICM
pressure. At the galaxy formation epoch the ICM pressure should 
have been much
smaller than this value.
  
In an expanding galactic wind flow the SN-heated gas will cool
adiabatically to quite small temperatures. However it will be reheated in
the termination shock, where the ram pressure of the wind adjusts to the
ICM pressure. Much beyond this point the ejected galactic gas is expected
to slowly exchange energy and some metal-rich material with the
unprocessed ICM gas.

\subsection{Nonthermal particle production}

\begin{figure}
\plotone{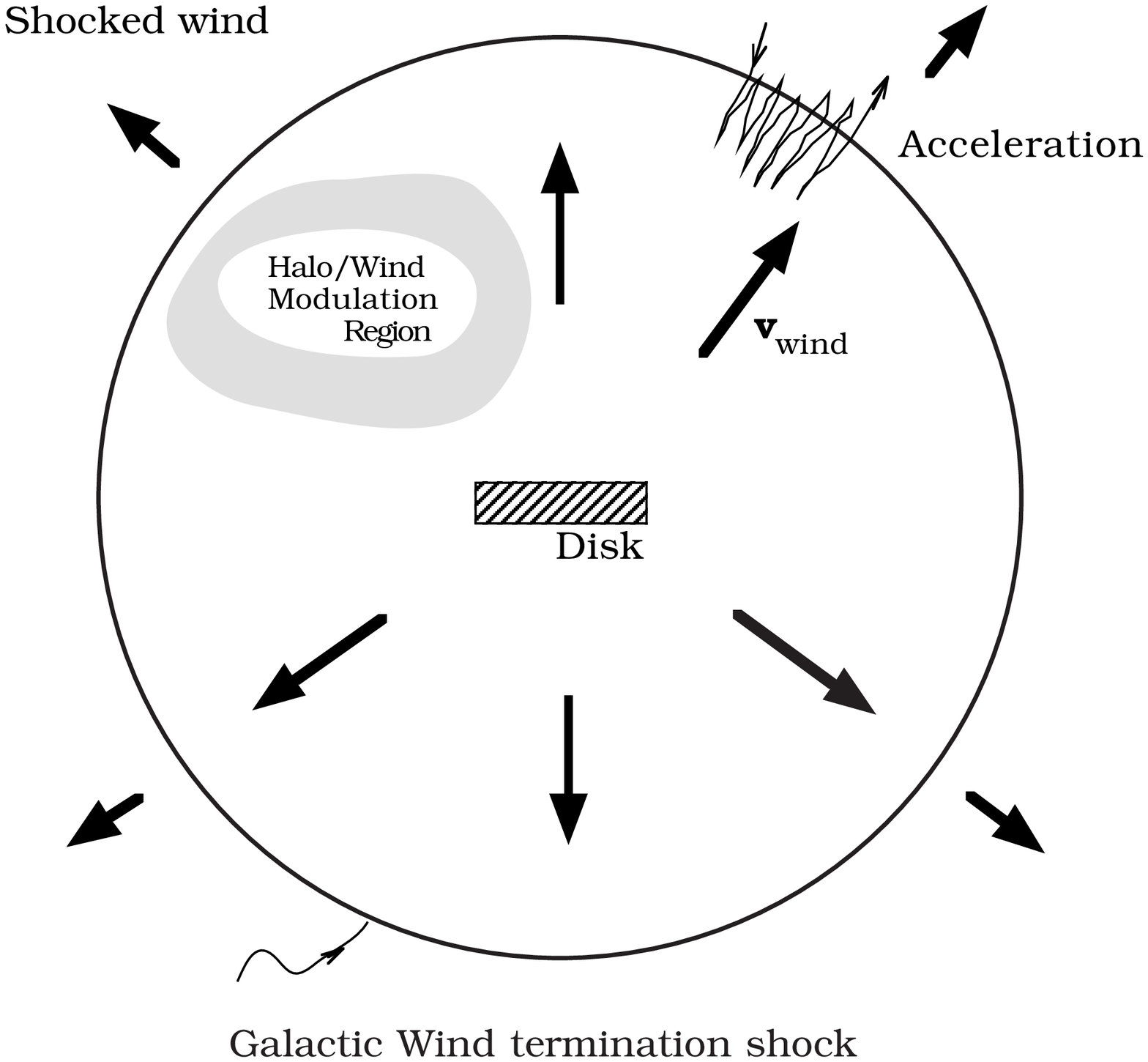} 
\caption{ 
 Schematic of particle acceleration at a galactic wind termination
shock. In the Halo, at distances large compared to the radius of the
galaxy (Disk), the wind velocity becomes radial and constant in magnitude.
It goes through a strong shock transition at a distance where the ram
pressure of the wind has decreased towards the external (ICM) pressure.
Particles injected from the heated thermal gas of the shocked wind get
diffusively accelerated at this shock with high efficiency, and slowly
convected into the ICM.
\label{fig1} }
\end{figure}

Cluster formation also implies the production of a strong nonthermal
component of relativistic particles. They will be accelerated during the
early phase - and presumably also in later accretion events that have
shock waves associated with them. In the Supernova Remnants (SNRs) the
main acceleration should occur at the outer shock, with a very high
efficiency of $\sim 10~{\rm percent}$ (\cite{dru89}; \cite{bervo2000}),
the
rest
of $E_{\rm SN}$ going into the thermal gas ($\sim 10~{\rm to}\, 20~ {\rm
percent}$) (\cite{dor93}), and radiation ($ \gsim 70 $\, percent).
However, since the particles are ultimately removed from the galaxies in a
strong galactic wind (see Paper I and section 3. below), they will cool
adiabatically like the thermal gas and transfer their energy to the
kinetic energy of the wind flow. As a consequence the original
SNR-accelerated particles should constitute a negligible fraction of the
present-day nonthermal particle content of the cluster. But this is not
the end of the story. At distances $\sim 100~{\rm kpc}$ from the galaxies,
fresh particle acceleration will occur at the strong galactic wind
termination shocks (Fig.~\ref{fig1}).  We estimate the overall
acceleration efficiency in these shocks to be again of the order of 10
percent\footnote{Note that this acceleration efficiency might as well be
lower, if the termination shocks were essentially perpendicular.}. Over
the early phase of galaxy formation, using our estimate for the total
number of SNe, this should result in a total gas internal energy in the
postshock region $E_{\rm gas}^{\rm GW} \sim {\rm few}\times 10^{62}\,$ erg
and a nonthermal energy $E_{\rm CR}^{\rm GW}\, \sim {\rm few}\times
10^{61}$ erg for a system like the Perseus cluster, ultimately driven by
star formation and the subsequent SN explosions. Since the confinement
time in the cluster (see next subsection) exceeds the cluster lifetime,
the energy spectrum of these particles in the ICM is basically the same as
the source spectrum at the termination shocks. In Cosmic Ray (CR) parlance
these are {\it cosmological} CRs.

\placefigure{fig1}

Since the galaxies are distributed across the cluster quasi-uniformly,
this should originally also be true for the nonthermal particle
population. The ensuing gravitational contraction/accretion of the cluster
will subsequently energize CRs and thermal gas at least adiabatically or,
more likely, shock accelerate/irreversibly heat both components so that
finally the total energy $E_{CR}$ of energetic particles should reach at
least the adiabatic value $\sim 3\times 10^{62}\, {\rm erg}\! \sim 1/30 \,
E_{\rm gas}$ in the cluster; $E_{\rm gas}$ now denotes the total present
internal energy of the ICM gas (Paper I).

It is worthwhile to compare the expected nonthermal energy with the
thermal energy content of the cluster galaxies. Assuming the stars
internally to be in virial equilibrium and, for purposes of estimate, all
of them to have a solar mass and radius, then $E_{\rm th}^{\rm star}\!  
\sim (3/10) GM_{\odot}^{2}/R_{\odot} \simeq 10^{48}\,\rm erg$. For a total
mass of about $10^{14} M_{\odot}$ contained in the {\it galaxies} of the
Perseus cluster this gives a total thermal energy in stars $ \sim\!
10^{62}\,\rm erg$, and thus $E_{\rm CR} \gsim \sum_{\rm gal}\sum_{\rm
stars} E_{\rm th}^{\rm star}$.  This means that the nonthermal ICM energy
should be at least as large as the total thermal energy content of all the
stars in all the galaxies together.

It has been argued more recently that, apart from star formation and
overall gravitational contraction, also individual giant radio galaxies
should have injected large and in fact comparable amounts of nonthermal
particles during the life time of a cluster (\cite{ens97,ber97}). This is
no doubt an important additional possibility. A weakness of this argument
consists in the fact that per se it is predicated on statistical knowledge
about the luminosity function for active galaxies in clusters in general,
and not on direct observations of the individual cluster to which it is
applied.

\subsection{Particle Confinement}

Energetic particle confinement has been discussed in a number of
papers in recent years (Paper I, Berezinsky et al. 1997,
Colafrancesco \& Blasi 1998). We review here this important point,
also in the light of the magnetic field structure to be discussed in
the next section.

The large-scale magnetic field in the ICM gas may be quite chaotic and not
well connected over distances exceeding typical intergalactic distances.
Thus energetic particles may not readily escape from the cluster due to
such topological characteristics. However, already a consideration of pure
pitch angle diffusion along straight magnetic field lines with superposed
turbulent fluctuations gives important insights into the confinement
properties of galaxy clusters. Standard quasilinear theory yields a
spatial diffusion coefficient $\kappa_{\parallel}$ along the large scale
field $B$ due to a power spectrum $P(k)$ of magnetic field
fluctuations with wavelength $\lambda=2 \pi /k$ in the following form:
$$\kappa_{\parallel}=
(1/3) c r_{\rm g}(p)
\frac{B^2}{\int_k^{\infty} {\rm d}k' P(k')}, $$
where $r_{\rm g}(p)$ is the gyro radius of a particle with momentum $p$,
$kr_{\rm g}(p)\simeq 1$, and $k$ denotes the wavenumber of the field
fluctuations. Let us assume a relative fluctuation field strength of order
unity at the inter-galaxy distance $1/k_{0}$, i.e. a totally turbulent
field $P(k_0) \times k_0 \sim B^2$ on this scale, and a power law
form of $P(k)=P(k_0) (k/k_0)^{-n}$. Then the diffusion time across the
cluster $T_{\rm esc} \sim R_{\rm cl}^2 / \kappa_{\parallel} > T_{H}$ for
$(cp)_{\rm protons} \lsim 10^{17}\, {\rm eV \; and} \lsim10^{15}\,{\rm
eV}$, for $n=3/2$ and $n=5/3$, respectively (Paper I). Also $t_{\rm
loss}^{\rm protons}\gg T_{H}$ for nuclear collisions in the ICM gas.  
Therefore - except at subrelativistic energies with their prevailing
Coulomb losses - up to these energies CR hadrons should accumulate in the
cluster since the galaxy formation epoch, and that is what we called {\it
cosmological} CRs before. The situation is different for relativistic
electrons, which suffer radiative losses. Electrons, and the nonthermal
radiation from galaxy clusters are discussed extensively in the paper by
Atoyan \& V\"olk (2000).

\subsection{Gamma-rays from cosmological CRs}

\begin{figure} 
\plotone{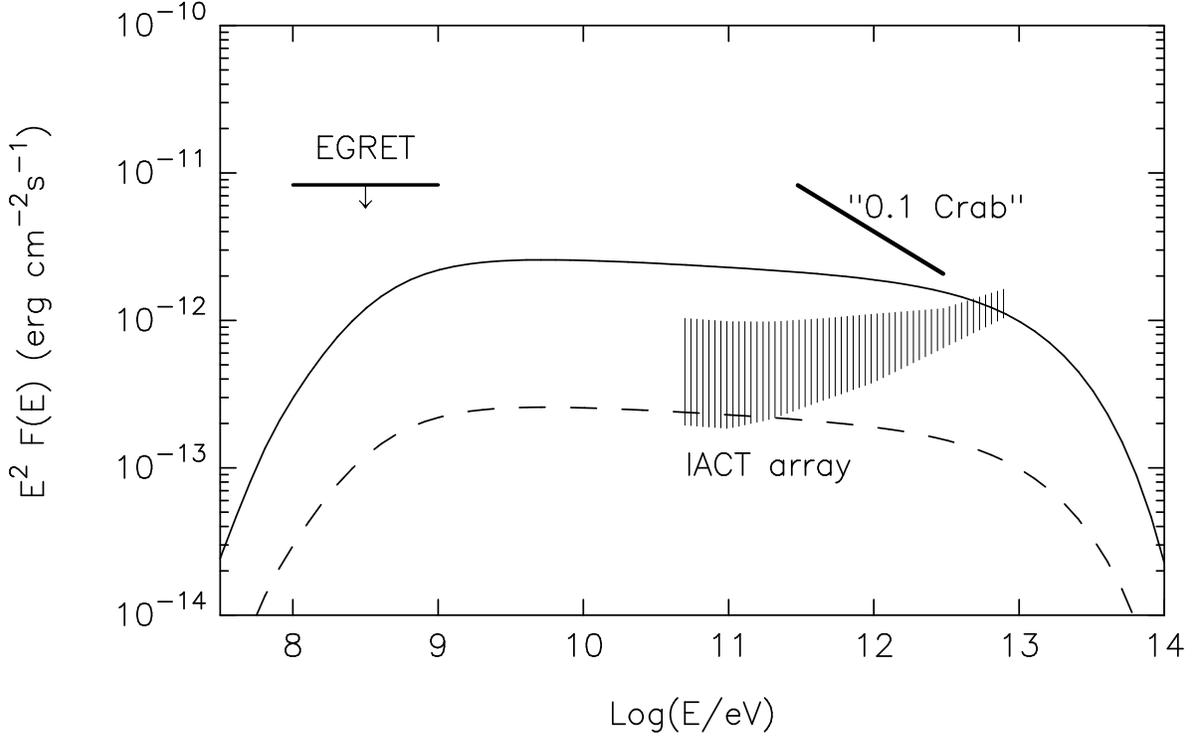}
\caption{Hadronic $\gamma$-ray fluxes expected from the Coma cluster for a
proton differential energy spectrum with spectral index $\alpha_{\rm CR}
=2.1$ and cutoff energy $E_{c}=200\, {\rm TeV}$.  The solid and the dashed
curve show the $\gamma$-ray fluxes produced in $pp$ interactions of CRs
with $E_{\rm CR}= 3\times 10^{62}\,\rm erg$ and $E_{\rm CR}= 3\times
10^{61}\,\rm erg$, respectively, in an ICM with $n=10^{-3}\,\rm cm^{-3}$;
the lower CR energy content might reflect a lower acceleration
efficiency at the GW termination shocks. Also
the EGRET upper limit is shown (Sreekumar et al. 1996). The heavy bar
shows the $10\,\%$ level of the TeV $\gamma$-ray flux from the Crab Nebula
(e.g. Konopelko et al. 1999). The light vertical bars show the limiting
fluxes for a 100\,h observation time with the H.E.S.S. imaging atmospheric
Cherenkov telescope (IACT) array of a 1 degree
extended source (upper ends) on the one hand, and of a point source (lower
ends) on the other (see Aharonian et al. 1997).
\label{fig2} }
\end{figure}

The accumulated cosmological CR protons and nuclei will produce high
energy $\gamma$ rays from inelastic pp-collisions on the Intracluster gas
which, in particular, lead to $\pi^0$-production and subsequent decay.
Fig.~\ref{fig2} shows the $\pi^0$-decay energy fluxes expected from the
hadronic CRs of the Coma cluster with a differential energy spectrum
$\propto E^{-{\alpha_{\rm CR}}}\,{\rm exp}(-E/E_{c})$, with $\alpha_{\rm
CR}=2.1$ and an upper cutoff energy $E_{c}=200\,{\rm TeV}$. The solid and
the dashed curve assume $E_{\rm CR}=3\times 10^{62}\,\rm erg$ and $E_{\rm
CR}=3\times 10^{61}$\, erg, respectively, in an ICM with a gas density of
$n=10^{-3}\,{\rm cm}^{-3}$. Also the EGRET upper limit by Sreekumar et al.
(\cite{sre96}) is shown. The observed size of the radio emission produced
by high-energy electrons in Coma is about half a degree. The brightness of
the {\it hadronic} TeV emission should be significantly higher in the
central region of the cluster, since the ICM gas density strongly
increases towards the center. Detection of an extended and weak
($\leq\,$0.1 Crab) TeV flux by current instruments is problematic, but
such fluxes are quite accessible for future imaging atmospheric Cherenkov
telescope arrays like H.E.S.S., VERITAS, and CANGAROO III (see Fig.\,2).

\placefigure{fig2}

\section{ Intracluster Magnetic Fields}

The magnetic field strengths in the ICM of rich clusters, which may be as
large as $B \sim 1 \,\mu \rm G$ as inferred by Faraday rotation
measurements (\cite{kro94}), are not easily explained by a contemporary
dissipative mechanism because present-day turbulent dynamo effects in such
a large-scale system should be extremely slow. This problem is compounded
by the extraordinary smallness of the expected intergalactic seed fields.
Therefore we suggest a field configuration that is due to the early
formation history of galaxy clusters as discussed above; it should be
preserved in its essential features to this day (\cite{volato99}). This
field should even be still in a state of development at the present epoch.
The model derives from the violent early galactic winds which
accompany the starbursts responsible for the predominance of the early
type galaxies in rich clusters.

In a general form the ejection of galactic fields has been
discussed by Kronberg (\cite{kro94}. For the generation of the general
Intergalactic magnetic field, related arguments have been advanced
independently by Kronberg et al. (\cite{kro99}), assuming very early
formation of dwarf galaxies with wind outflows at redshifts $\geq 10$.

We assume first of all that the gas-rich progenitors whose mergers
supposedly constitute the building blocks for the E and S0 galaxies, can
be specifically pictured as protospirals that had already generated
galactic magnetic fields of $\mu$G strength. This should indeed be
possible within a time of $10^8$ yr or less, i.e. a time scale of the
order of a rotation period of our Galaxy or even shorter. The explanation
derives from fast turbulent dynamo action that invokes boyancy effects due
to CRs that inflate the magnetic flux tubes together with magnetic
reconnection over spatial scales of order $100 {\rm pc}$ (\cite{par92}).
In starburst galaxies also systematic dynamo effects might play an
important role (section 3.1).

In a second stage, i.e. during the galaxy mergers, the resulting
supersonic galactic winds will extend these fields from the interacting
galaxies to almost intergalactic distances. In the final and by far
longest stage that lasts until now, the fields should be recompressed by
the contraction of the cluster to its present size\footnote{For a
different view, emphasizing internal shear flows, see Dolag et al.
(\cite{dol99}.)}. In addition, the continuing accretion of subclusters and
individual galaxies constantly perturbs this field, keeping the
fluctuations around this large scale field at a high level.

The ICM fields do not reconnect on the intergalactic scale in a Hubble
time. Consequently there is no need for a continuous regeneration of these
fields since their formation. However, this also implies that a
topologically connected overall ICM field will on average not be formed
either and that the ICM field is chaotic on a scale smaller or equal to
the present intergalactic distance.

\begin{figure}
\plotone{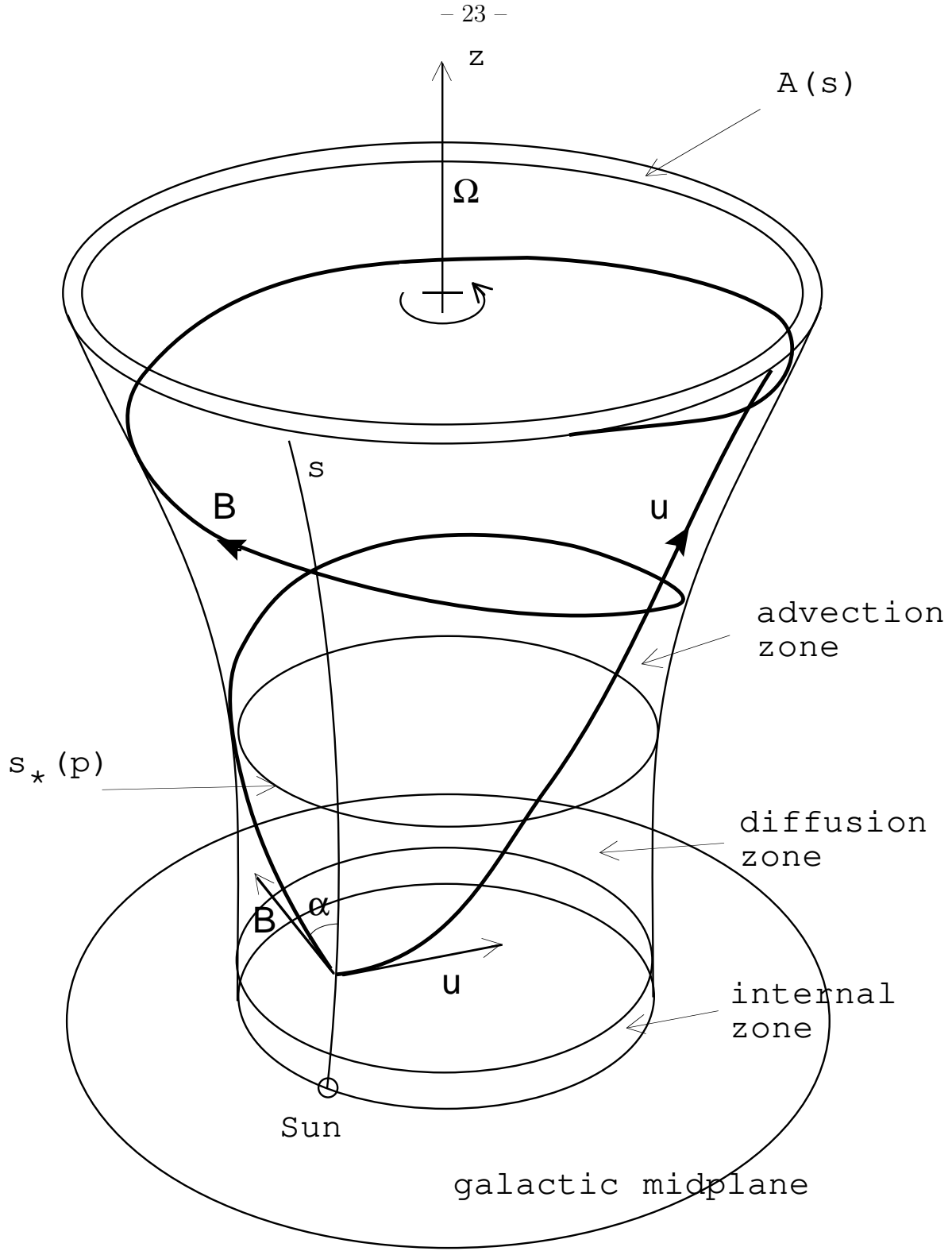}
\caption{
Schematic of the flow and field configuration of an 
axisymmetric Galactic Wind from a rotating disk galaxy like the Milky Way, 
cf. Ptuskin et al. (1997). 
The meridional flow velocity {\bf u} from a circle above the
disk (like that corresponding to the Sun's distance from the Galactic
Center) is roughly parallel to the axis of rotation, to flare out
at
large distances. The magnetic field $B$ becomes slowly azimuthal at
large meridional distances $s$.  
\label{fig3}}
\end{figure}

In detail we draw on arguments which we have used in the past for the
field configuration in a galactic wind from our own Galaxy
(\cite{zir96,ptu97}), see also Fig. 3. They are based on estimates of the
relative amount of field line reconnection vs. the extension of Galactic
field lines by a wind to "infinity" (\cite{bre93}). The basic result is
that the rates of reconnection - and thus of the formation of "Parker
bubbles" leaving the Galaxy by their boyancy and allowing the generation
of the disk magnetic field - and the rates of extension of this field into
the Galactic Halo by the pressure forces of the wind are roughly equal.
Thus both effects occur with about equal probability. For the cluster
galaxies this means that magnetic energy can be generated on the large
scale of the wind at the expense of the thermal and nonthermal enthalpies
produced in the starburst. The geometry of the field should roughly
correspond to straight field lines out to meridional distances $s$ of the
order of the starburst (SB) radius, $R_{\rm gal}^{\rm SB} \sim 1$ kpc in
the protogalactic disk, and spherically diverging field lines beyond that.  
The slow rotation of the system should then lead to an azimuthal field
component, decreasing with the wind distance $\propto 1/s$, which
dominates at large distances over the radial component. However, in
contrast to the familiar situation in the Solar Wind equatorial plane, the
axis of rotation is rather parallel than perpendicular to the flow at the
base of the wind, and thus the dominance of the azimuthal field component
is by no means as drastic as in the case of a stellar wind
(Fig.~\ref{fig3}). The wind becomes supersonic at about the same critical
distance, $s_{crit} \sim R_{\rm gal}^{\rm SB}$. Far beyond this critical
point the mass velocity $u(s)$ becomes constant and the density falls off
$\propto s^{-2}$.
\placefigure{fig3}
\subsection{Mean magnetic field}

Choosing a present average baryon density $n_{\rm b}(z=0)=3 \times
10^{-7}{\rm
cm}^{-3}$, i.e. assuming most of the baryonic matter to be in the form of
Intergalactic gas, the mean density at the formation stage of the early
type galaxies (at redshifts $z\simeq 2$) was then $n_{\rm b}(z=2) = 27
n_{\rm b}(z=0)\equiv n_{\rm cl}(z=2)$. With a present ICM density
$n_{\rm cl}\gsim~
10^{-4}{\rm cm}^{-3}$, we have $n_{\rm cl}(z=0)/n_{\rm cl}(z=2)\gsim~ 12$
due to
gravitational compression of the ICM gas. With a dominant thermal gas
pressure the corresponding adiabatic pressure increase $p_{\rm cl}\propto
n^{5/3}$ amounts to $p_{\rm cl}(z=0)/p_{\rm cl}(z=2) \gsim~ 63$; for the
following
we shall use $p_{\rm cl}(z=0)/p_{\rm cl}(z=2) = 10^2$. The wind
termination
shock distance $s_{\rm sh}$ is then given by $\rho(r_{\rm sh})\, u^2 \sim
p_{\rm cl}(z=2)$, where $\rho(r_{\rm sh})$ is the wind mass density
upstream of the shock.

To estimate the wind characteristics we assume a quasi-steady state and a
strong starburst for which we may disregard gravity, magnetic forces and
CR pressure gradients in the overall energy balance equation (e.g.
\cite{zir96}). It reads in this case:

$$
\left[\frac{u^2}{2}+\frac{\gamma}{\gamma-1}\frac{p}{\rho}\right]_s
\simeq \frac{\gamma}{\gamma -1}\frac {p_{\rm base}}{\rho_{\rm base}},
$$

\noindent where in addition $\rho u^2(s)/2 \ll \gamma p/(\gamma -1)$ is
assumed at
the base $s=s_{\rm base}$ of the wind. The critical point $s_{\rm crit}$
is given
by $\rho_{\rm crit}u_{\rm crit}^2=\gamma p_{\rm crit}$. Approximately
$u_{\rm crit} \simeq
u_{\infty}/2$, where $u_{\infty}$ is the asymptotic wind speed. Beyond the
critical point at $s_{\rm crit}\simeq R^{\rm SB}_{\rm gal}$ the wind
achieves spherical symmetry so that from mass conservation

$$
\left(\frac{s_{\rm sh}}{R^{\rm SB}_{\rm gal}}\right)^2 \simeq
\frac{\rho_{\rm crit}u_{\rm crit}}{\rho_{\rm sh} u_{\rm sh}}
$$

If the ICM pressure is small compared to the base pressure then
we can
assume that the wind is highly supersonic at the shock distance which
implies $u_{\rm sh} \simeq u_{\infty}$. This should be true for M82-like
objects - although they should have a considerably larger scale
$R_{\rm gal}^{\rm SB} \sim
1$ kpc - which have $p_{\rm gas}/k_{\rm B} \sim 10^7 \,{\rm K}{\rm
cm}^{-3}$.
In this case we can assume the wind pressure to evolve adiabatically to
lowest order, so that

$$
\frac{p_{\rm crit}}{p_{\rm base}} \simeq
\left(\frac{\rho_{\rm crit}}{\rho_{\rm base}}\right)^{5/3}
$$
As a result, taking into account that in a flow dominated by the thermal
gas the adiabatic index $\gamma=5/3$
$$
\frac{s_{\rm sh}}{R^{\rm SB}_{\rm gal}} \simeq \left[\frac{4\gamma}{\gamma
+1}\left(\frac{1}{2(\gamma -1)}\right)^{\frac{1}{\gamma
-1}}\right]^{1/2} \left(\frac{p_{\rm base}}{p_{\rm cl}}\right)^{1/2}
\simeq
1.27~
\left(\frac{p_{\rm base}}{p_{\rm cl}}\right)^{1/2} \simeq 400 
$$
for the above starburst parameters.

Therefore 
$$ 
\frac{s_{\rm sh}}{d_{\rm gal}^{\rm field}(0)/(1+z)} \simeq \frac{400
\,{\rm kpc}}{(2\,{\rm Mpc}/3)} \simeq 0.6
$$ 

The Wind Bubble containing the shock-heated wind gas will have a radius
still exceeding $s_{\rm sh}$. Beyond the Bubble, part of the external gas
will be shocked by the rapidly expanding Bubble gas. This shock heated gas
will exchange energy with the cold ambient gas by heat conduction and
instabilities. To some extent such exchanges will also take place with the
Bubble gas. 

Even though the volume of at least initially unmagnetized ICM gas may be
large enough so that the ICM gas mass exceeds the mass associated with
galaxies by a factor of a few - as observed - it may therefore be that
some Wind Bubbles touch. Thus we should consider whether the magnetic
fields of the bubbles can reconnect with each other to produce field lines
that pervade the entire cluster. Let us assume that the fastest rate of
reconnection proceeds with a speed between 1 and 10 percent of the
Alfv$\acute{\rm e}$n speed $v_{\rm A}$, to be specific, with $v_{\rm
A}/50$ (\cite{par92}). Then the present-day reconnection time across a
termination shock scale is
$$
t_{\rm rec} \sim \frac{s_{\rm sh}}{(v_{\rm A}/50)} \simeq 9\times 10^{10}
{\rm yr}
\left\{\left(\frac{s_{\rm sh}}{400 {\rm kpc}}\right)\left(\frac
{B}{10^{-6}{\rm G}}\right)^{-1}
\left(\frac{n}{10^{-4}}\right)^{1/2}\right\},
$$
greater than a Hubble time, even for a present-day ICM
magnetic field as high as $10^{-6}$ G. Therefore, many bubble fields
might not yet be reconnected and much of the field structure could well
remain topologically disconnected until today.

The field strength $B_{\rm cl}(z=2)$ in the early Wind Bubbles
should be
of the order of
$$
B_{\rm cl}(z=2) \simeq 4 B_{\rm gal} r_{\rm gal}^{\rm
SB}/r_{\rm sh}
\sim
10^{-2}B_{\rm gal} \sim 10^{-8} {\rm G}
$$
or somewhat larger, if the field in the bubble increases in the
decelerating postshock flow.
The subsequent and still ongoing overall cluster
contraction/accretion compresses the field to lowest
order isotropically $\propto l^2$, with the scale factor
$$
l \simeq [n_{\rm cl}(0)/n_{\rm bar}(0)]^{1/3}/(1+z),
$$
where $n_{\rm cl}(0)\sim (10^{-3}~{\rm to}~
10^{-4})\,{\rm cm}^{-3}$ and $z=2$.
For a present mean baryon number density 
$n_{\rm bar}(0) \sim 3 \times 10^{-7}\,{\rm cm}^{-3}$, we obtain $l \sim 
(2.3~{\rm to}~ 5)$. 

From these estimates, we obtain for the present-day ICM field:
$B_{\rm cl}(z=0)/B_{\rm cl}(z=2) \sim 5 ~{\rm to} ~25$.
Therefore the
present-day ICM magnetic field should have a mean strength of the order of
$10^{-7}$G, from $1 {\mu}{\rm G}$ "primordial" seed fields, and should be
randomly directed on an intergalactic scale. Although smaller by about one
order of magnitude than estimated from Faraday rotation measurements, such
fields need not necessarily be unrealistic, considering that observations
might emphasize regions of high magnetic fields. Indeed the simplest
inverse Compton interpretation of the EUV excess and the excess X-ray flux
in the Coma cluster requires such low field strengths (\cite{fus99}).  

On the other hand, the increase of the field in the galactic Wind Bubbles
beyond their postshock value might be more than a factor of unity as
assumed above. 

An additional possibility is that the "initial" fields for
such starburst galaxies might be an order of magnitude stronger than
assumed. Apart from a fast dynamo whose strength could be directly
proportional to the star formation rate (\cite{par92}), we could invoke a
systematic field amplification through the commencing galactic outflow
that "combs" the field outward. It might occur on very short time scales
of about $10^7$ yr. Empirically this field amplification is suggested by 
the statistical time-independence of the radio synchrotron to
far infrared emission ratio in starburst galaxies (\cite{lis96}) which can
hardly be understood otherwise than through a field that increase almost
simultaneously in strength with the star formation rate.

A final argument is that the large scale shear deformations induced by
the later accretion of large subclusters may amplify the field even
further. Thus we cannot exclude $\mu$G fields; the force balance
certainly allows them.

\subsection{Field evolution and structure}

The Wind Bubbles and the associated magnetic fields should at some stage
decouple from the galaxies they emanated from, simply by magnetic
reconnection which is fastest near the galaxies: the Alfv$\acute{\rm e}$n
velocity is approximately independent of the meridional distance s in the
wind, whereas the distance between oppositely directed field lines that
emanate from the galaxy is obviously the smaller the smaller s is. Thus,
after the termination of the starburst, the magnetized bubbles and the
stellar component of the remaining early type galaxies should acquire
independent identities and their dynamics should decouple.

The topologically disconnected structure of the mean magnetic field in the 
cluster has also some bearing on the evolution of magnetic field strengths
during the development of cooling flows towards the cluster center (\cite
{fab94}). Instead of building up global magnetic pressure gradients and
tension forces such a subsonic and sub-Alfv$\acute{\rm e}$nic
flow allows optimal internal segregation of high and low field regions on 
galaxy - galaxy seperation scales. Thus we should expect that the field
strength towards the cluster center increases by less than by isotropic 
compression, $B \propto n^{2/3}$, even though at the compressed spatial
scales reconnection will be more effective. The two-thirds law should hold
only for the overall cluster gas compression discussed earlier.

A question of direct importance for the interpretation of Faraday rotation
measurements is the scale of reversal changes of these fields. Field
reversals are a natural consequence of the suggested field structure,
given the field pattern in the interstellar media of the starburst
galaxies. If the Milky Way can serve as a guide, then this pattern is
determined by the 100 pc scale of the Parker instability (Parker 1966).
From the radius $r_{\rm gal}^{\rm SB}$ to distances $\sim r_{\rm sh}$ this
scale projects like the ratio $r_{\rm sh}/ r_{\rm gal}^{\rm SB}$. Using
the above estimates this implies a field reversal scale $\simeq 40$\, kpc
in clusters, rather well in line with observational estimates which
indicate reversal scales of $10$ to $100$ kpc (\cite{kro94};
\cite{clarke99}).

\subsection{Magnetic field fluctuations} 

From the foregoing arguments there is hardly any need for a contemporary
"turbulent ICM dynamo". However, the ongoing accretion will perturb this
mean intracluster field randomly in space and time, maintaining a
turbulent magnetic fluctuation field that develops smaller and smaller
spatial scales. Due to the topology of the mean field, the largest
turbulent scale is given by the distance between galaxies. The accretion
will probably span the range from single field galaxies falling into the
cluster to accreting massive subclusters. As long as this accretion
process remains important, one has to expect that the cluster rings with
it. At the largest scale the relative magnetic fluctuation level should
approach unity (see subsection 2.3).

\begin{figure}
\plotone{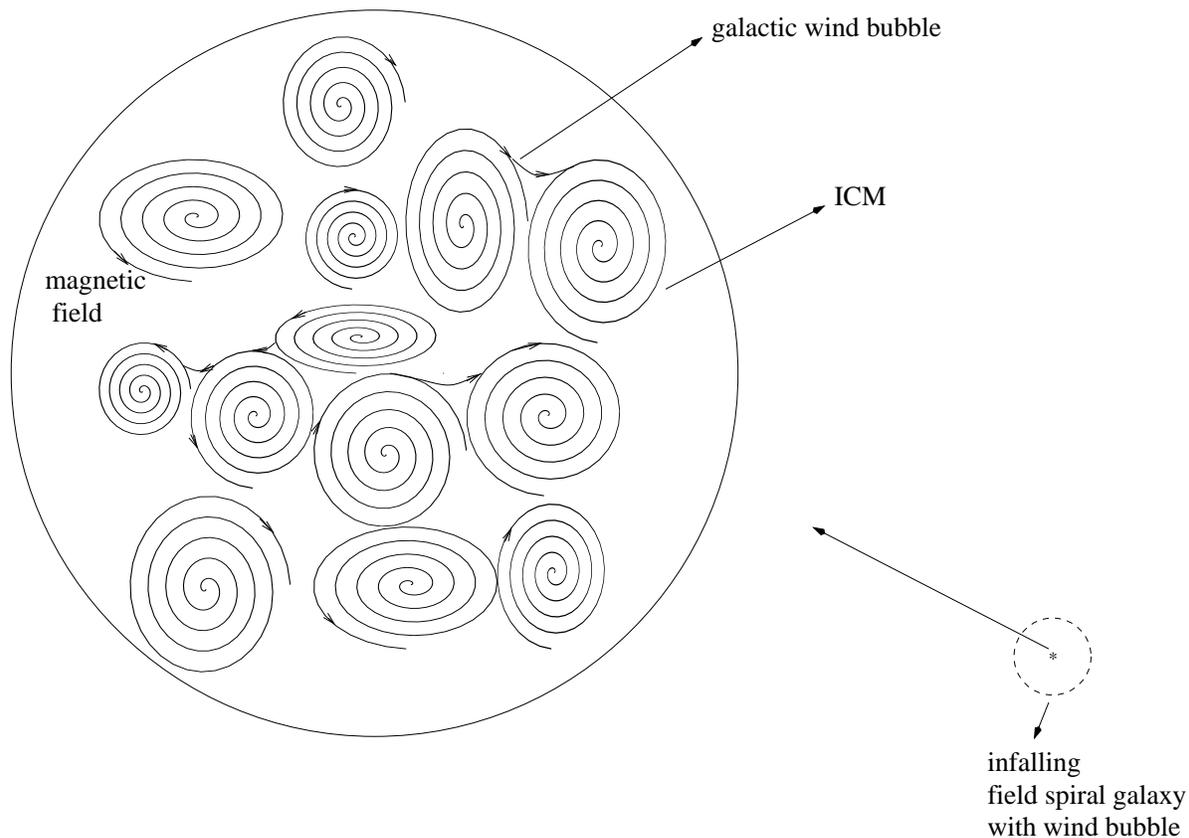}
\caption{Cartoon of the large-scale magnetic field topology in a
cluster.
The magnetic fields have been originally generated in the (rotating)
gas-rich galaxies and are then drawn out by strong galactic winds. The
fields are frozen into an ensemble of randomly oriented bubbles created by
the winds from early starbursts. For demonstration purposes less than 10
percent of all the bubbles expected are shown. Due to reconnection in
their interstellar media the present-day galaxies are no more dynamically
connected to the bubbles. However magnetic reconnection is rather
ineffective in the ICM at large. Therefore only some of the bubbles will
be magnetically connected to their neighbors. Other bubbles just contain
hot gas and closed magnetic field loops. The bubbles are surrounded by
less magnetized Intracluster gas that contains much of the cluster mass.
However they exchange energy by thermal conduction, and potentially by
instabilities. Cooling flows are not considered in the Figure, but
possible, and in fact likely to occur. The configuration is continuously
perturbed by the infall of single field galaxies and the accretion of
subclusters
that lead to shear amplification and distortion of the simple idealized
structures shown, as well as to a high level of magnetic fluctuations.
\label{fig4} }
\end{figure}

\placefigure{fig4}

 \acknowledgements 
{The authors thank
F.\,A.\,Aharonian, E.\,N.\,Parker, and R.\,Rosner for illuminating
discussions.  The work of AMA was supported through the
Verbundforschung Astronomie/Astrophysik of the German BMBF under the
grant No. 05-2HD66A(7).}


\clearpage

\begin{thebibliography}{}
%
\bibitem{} 
Aharonian, F. A., Hofmann, W., Konopelko, A. K., \& V\"olk, H. J.
1997, Astropart. Phys., 6, 343; 369
%
\bibitem[Arnaud et al. 1992]{arn92} Arnaud, M., Rothenflug, R., Boulade, O., 
 Vigroux, L., \& Vangioni-Flam, E. 1992, A\&A,  254, 49
%
\bibitem[Atoyan \& V\"olk 1999]{atovol2000} Atoyan, A. M., \& V\"olk, H.
J. 1999, ApJ in the press
%
\bibitem[Berezhko \& V\"olk 2000]{bervo2000} Berezhko, E.G., \& V\"olk,
H.J. 2000, ApJ submitted
%
\bibitem[Berezinsky et al. 1997]{ber97} 
Berezinsky, V. S., Blasi, P., \& Ptuskin, V. S. 1997, \apj,
487, 529
%
\bibitem[e.g. B\"ohringer 1996]{boh96}  B\"ohringer, H. 1996,
Extragalactic Radio Sources, R. Ekers et al., 1996, IAU, 357
%
\bibitem[Breitschwerdt et al. 1993]{bre93} Breitschwerdt D., 
McKenzie J. F., \& V\"olk H. J. 1993, A\&A, 269, 54
%
\bibitem[Breitschwerdt 1994]{bre94} Breitschwerdt D., 1994, 
Habilitationschrift, Univ. Heidelberg, 158
%
bibitem[Clarke et al. 1999]{clarke99} Clarke, T.E., Kronberg, P.P.,
B\"ohringer, H. 1999, Proc. of the Ringberg Workshop: "Diffuse Thermal and
Relativistic Plasma in Galaxy Clusters", Ringberg Castle, Germany, April
19-23, 1999, eds. H. B\"ohringer, L. Ferretti \& P. Sch\"ucker, MPE Report
271, 82
%
\bibitem[Colafrancesco \& Blasi 1998]{colbla98} Colafrancesco, S., \& Blasi, P. 
1998, Astropart. Phys. 9, 227
%
\bibitem[1998]{dol99} Dolag, K. M., Bartelmann, M., \& Lesch, H. 
1999, A\&A 348, 351
%
\bibitem[Dorfi 1993]{dor93} Dorfi, E. A. 1993, Galactic High-Energy Astrophysics,
High-Accuracy Timing and Positional Astronomy, J. van Paradijs and H. M.
Maitzen, Lecture Notes in Physics 418, Berlin: Springer, 1993, 43 
%
\bibitem[Drury et al. 1989]{dru89} Drury, L. O'C., Markiewicz, W. J., 
\& V\"olk, H. J. 1989, A\&A, 225, 179
%
\bibitem[En{\ss}lin et al. 1997]{ens97} En{\ss}lin, T. A., Biermann P. L., 
Kronberg, P. P., Wu, X.-P. 1997, \apj, 477, 560
%
\bibitem[Fabian 1994]{fab94} Fabian, A.C. 1994, ARAA 32, 277
%
\bibitem[Fusco-Femiano et al. 1999]{fus99} Fusco-Femiano, R., Dal Fiume, D., 
Feretti, L., Giovannini, G., Grandi, P., Matt, G., Molendi, S., 
\& Santangelo, A. 1999, \apj, 513, L21

\bibitem[see e.g. Kauffmann et al. 1999]{kau99} 
Kauffmann, G., Colberg, J. M., Diaferio, A., \& White, S. D. M. 1999,
\mnras, 303, 188

\bibitem{kon99} Konopelko, A. K., for the HEGRA 
collaboration. 1999, Rayos Cosmicos 98: Proc. 16th European Cosmic Ray
Simposium, Madrid, J. Medina;  (preprint astro-ph/9901094)

\bibitem[e.g. Kronberg 1994]{kro94} Kronberg, P. P. 1994, 
Rep. Prog. Phys., 57, 325. 

\bibitem[1999]{kro99} Kronberg, P. P., Lesch, H., \& Hopp, U. 
1999, \apj, 511, 56.

\bibitem[Lisenfeld et al. 1996]{lis96} Lisenfeld, U., V\"olk, H.J., \& Xu,
C. 1996, A\&A 314, 745

\bibitem[Parker 1966]{par66} Parker, E. N. 1966, \apj, 145, 811

\bibitem[Parker 1992]{par92} Parker, E. N. 1992, \apj, 401, 137

\bibitem[Ptuskin et al. 1997]{ptu97} Ptuskin, V. S., V\"olk, H. J., 
Zirakashvili, V. N., \& Breitschwerdt, D. 1997, A\&A, 321, 434

\bibitem[Schaaf et al. 1989]{sch89} Schaaf, R., Pietsch, W., Biermann, P. L., 
Kronberg, P. P., \& Schmutzler, T. 1989, ApJ, 336, 722

\bibitem[1996]{sre96} Sreekumar, P., et al. 1996, 
ApJ, 464, 628

\bibitem[Steinmetz 1993]{ste93} Steinmetz, M. 1993, PhD thesis, 
Technische Universit\"at M\"unchen

\bibitem[Turner 1999]{tur99} Turner, M. S. 1999, to appear in 
Physica Scripta; preprint astro-ph/9901109


\bibitem[1996]{paper1} V\"olk, H. J., Aharonian, F. A., 
\& Breitschwerdt, D. 1996, Space Sci. Rev., 75, 279; (Paper I)

\bibitem[V\"olk \& Atoyan 1999]{volato99} V\"olk, H. J., \& Atoyan, A. M. 1999, 
Astropart. Phys., 11, 73 

\bibitem[e.g. White \& Fabian 1995]{whifab95} White, D. A., \& Fabian, A. C. 
1995, \mnras, 273, 72

\bibitem[Zirakashvili et al. 1996]{zir96} Zirakashvili, V. N., Breitschwerdt, D.,
Ptuskin, V. S., \& V\"olk, H. J. 1996, A\&A, 311, 113

\end{thebibliography}
\end{document}